\documentstyle[prd,aps,floats,graphicx]{revtex}

\begin{document}
\preprint{ astro-ph/0008318}
\draft

%
%
\input epsf
\renewcommand{\topfraction}{0.8}
\twocolumn[\hsize\textwidth\columnwidth\hsize\csname
@twocolumnfalse\endcsname

\title{The WIMP annual modulation signal and non-standard halo models}
\author{Anne M.~Green} 
\address{Astronomy Unit, School of Mathematical
Sciences, Queen Mary and Westfield College, Mile End Road, London, E1
4NS, UK} 
\date{\today} 
\maketitle

\begin{abstract}
Currently the best prospect for detecting Weakly Interacting Massive
Particles (WIMPs) is via the annual modulation, which occurs due to
the Earth's rotation around the Sun, of the direct detection
signal. We investigate the effect of uncertainties in our knowledge of
the structure of the galactic halo on the WIMP annual modulation
signal. We compare the signal for three non-standard halo models:
Evans' power-law halos, Michie models with an asymmetric velocity
distribution and Maxwellian halos with bulk rotation. We then compare
the theoretical predictions of these models with the experimental
signal found by the DAMA experiment and investigate how the WIMP mass
and interaction cross section determined depend on the halo model
assumed. We find that the WIMP mass confidence limits are
significantly extended to larger masses, with the shape of the allowed
region in the mass-cross section plane depending on the model.
\end{abstract}

\pacs{98.70.V, 98.80.C }

\vskip2pc]

\section{Introduction}
The rotation curves of spiral galaxies are typically flat out to about
$\sim 30$ kpc. This implies that the mass enclosed increases linearly
with radius, with a halo of dark matter extending beyond the luminous
matter~\cite{ash}. The nature of the dark matter is
unknown~\cite{pss}, with possible candidates including MAssive
Compact Halo Objects (MACHOs), such as brown dwarves,
Jupiters or black holes and elementary particles, known as Weakly
Interacting Massive Particles (WIMPs), such as axions and the
neutralino, the lightest supersymmetric particle.

WIMPs can be directly detected via their elastic scattering off target
nuclei. In the long term the directional dependence of the detector
recoil will provide the best means of direct WIMP
detection~\cite{ck}. Currently the best prospect for distinguishing a
WIMP signal from the detector background is via the annual modulation
of the signal, which occurs due to the Earth's rotation around the
sun~\cite{am1,am2}.  The DAMA collaboration~\cite{dama}, using a
detector consisting of radiopure NaI crystal scintillators at the Gran
Sasso Laboratory, have recently reported a $4 \sigma$ annual
modulation signal consistent with the detection of a WIMP with mass
$m_{\chi}= 52^{+10}_{-8}$ GeV~\cite{newdama}, assuming a Maxwellian
halo with velocity dispersion roughly equal to the local rotation
velocity, $v_{0}=220 \rm km s^{-1}$ .

The values of the WIMP mass and interaction cross section found from
the annual modulation signal are known to depend somewhat on the
assumed values of poorly known astrophysical parameters. The effects
of bulk rotation~\cite{rot1,rot2} and varying the velocity
dispersion~\cite{br,rot2}, within the observationally allowed range
$v_{0}=220 \pm 40 \rm km s^{-1}$~\cite{vrange}, have been examined for
a Maxwellian halo. The DAMA collaboration have subsequently included
these uncertainties in the analysis of their latest
data~\cite{newdama}.

The standard Maxwellian halo model has a number of deficiencies (for
example see Ref.~\cite{alcock95} and references therein). The halo may
not be spherical; N body simulations of gravitational collapse produce
axisymmetric or triaxial halos, and indeed other spiral galaxies
appear to have flattened halos~\cite{Sackett}. The power-law halo
models of Evans~\cite{evans} provide an analytically tractable
framework for investigating the effect of varying halo properties such
as the flattening, and have previously been used to investigate
variations in the mean total~\cite{kk} and directional~\cite{ck} WIMP
detection rates.  Another possible modification to the standard halo
model is an asymmetric velocity distribution~\cite{am1} and the annual
modulation of the total~\cite{am1,Vergados,UK} and
directional~\cite{Vergados} WIMP signals have been calculated for
various asymmetric velocity distributions.

In this paper we compare the variation of the annual modulation
signal for power-law halo models, asymmetric velocity distributions
and Maxwellian halos with bulk rotation. We then examine how the range
of WIMP masses consistent with the latest DAMA data depends on the
halo model assumed.

\section{Annual Modulation signal}
The WIMP detection rate depends on the speed distribution of the WIMPs
in the rest frame of the detector, $f_{v}$. This is found from the
halo velocity distribution, $f({\bf v})$ by making a Galilean
transformation ${{\bf v}} \rightarrow {\tilde{\bf v}}= {\bf v} + {\bf
v_{{\rm e}}}$, where ${\bf v_{{\rm e}}}$ is the Earth's velocity
relative to the galactic rest frame, and then integrating over the
angular distribution~\footnote{The effect of the Sun's gravity on the
WIMP distribution can be neglected~\cite{griest}.}  In galactic
co-ordinates the axis of the ecliptic lies very close to the $\phi-z$
plane and is inclined at an angle $\gamma \approx 29.80^{\circ}$ to
the $\phi-r$ plane~\cite{Vergados}. Including all components of the
Earth's motion, not just that parallel to the galactic rotation:
\begin{eqnarray}
{\bf v_{{\rm e}}}& =& v_{1} \sin \alpha \, {\hat r}  +  \nonumber \\
                 &&    (v_{0} +v_{1} \cos \alpha
                    \sin \gamma ) \,  {\hat \phi} -  v_{1} \cos \alpha
                   \cos \gamma  \, {\hat z}\,,   
\end{eqnarray}   
where $v_{0} \approx 232 {\rm km s^{-1}}$ is the speed of the sun with
respect to the galactic rest frame, $v_{1} \approx 30 {\rm km s^{-1}}
$ is the orbital speed of the Earth around the Sun and 
$\alpha= 2 \pi (t-t_{{\rm
0}})/ T$, with $T=1$ year and $t_{{\rm 0}} \sim 153$ days (June 2nd), 
when the component of the Earth's velocity parallel to the Sun's motion 
is largest.

In the range of masses and interaction cross sections accessible to
current direct detection experiments the best motivated WIMP candidate
is the neutralino, for which the event rate is dominated by the scalar
contribution.  The differential event rate can then be written in
terms of the WIMP cross section on the proton, 
\begin{equation}
\sigma_{{\rm p}}= \frac{ 4 m_{{\rm p}}^2 m_{\chi}^2}{ \pi (m_{{\rm p}}
                + m_{\chi})^2 } f_{{\rm p}}^2 \,,
\end{equation} 
where $m_{{\rm p}}$ is the proton mass and $f_{{\rm p}}$ the effective
WIMP cross section on the proton.
The differential event rate simplifies to (see the Appendix for
more details):
\begin{equation}
\frac{{\rm d} R}{{\rm d}E} = \xi \sigma_{{\rm p}} 
              \left[ \frac{\rho_{0.3}}{\sqrt{\pi} v_{0}}
             \frac{ (m_{{\rm p}}+ m_{\chi})^2}{m_{{\rm p}}^2 m_{\chi}^3}
             A^2 T(E) F^2(q) \right] \,,
\end{equation}
where the local WIMP density, $\rho_{\chi}$ has been normalised to a
fiducial value $\rho_{0.3} =0.3 {\rm GeV/ cm^{3}}$, such that
$\xi=\rho_{\chi} / \rho_{0.3}$, $E$ is the recoil energy of the
detector nucleus and $T(E)$ is defined as~\cite{jkg}
\begin{equation}
\label{tq}
T(E)=\frac{\sqrt{\pi} v_{0}}{2} \int^{\infty}_{v_{{\rm min}}} 
            \frac{f_{v}}{v} {\rm d}v \,,
\end{equation}
where   $v_{{\rm
min}}$ is the minimum detectable WIMP velocity
\begin{equation}
v_{{\rm min}}=\left( \frac{ E (m_{\chi}+m_{A})^2}{2 m_{\chi}^2 m_{A}} 
             \right)^{1/2} \,,
\end{equation}
and $m_{A}$ the atomic mass of the target nuclei.

In order to compare the theoretical signal with that observed we need
to take into account the response of the detector. The electron
equivalent energy, $E_{{\rm ee}}$, which is actually measured is a
fixed fraction of the recoil energy: $E_{{\rm ee}}= q_{{\rm A}}
E$. The quenching factors for I and Na are $ q_{{\rm I}}=0.09$ and
$q_{{\rm Na}}=0.30$ respectively~\cite{q}. The energy resolution of
the detector~\cite{rot1} is already taken into account in the data
released by the DAMA collaboration.

The expected experimental spectrum per energy bin for the DAMA
collaboration set-up is then given by~\cite{br}
\begin{eqnarray}
\frac{\Delta R}{\Delta E} (E) &=& r_{{\rm Na}} \int_{E/q_{{\rm
             na}}}^{(E+ \Delta E)/q_{{\rm na}}} \frac{{\rm d}R_{{\rm
             Na}}  }{{\rm d} E_{{\rm ee}}} (E_{{\rm ee}})  \frac{{\rm d}
             E_{{\rm ee}}}{ \Delta E} \nonumber \\ &&  + \, r_{{\rm I}}
              \int_{E/q_{{\rm
             I}}}^{(E+ \Delta E)/q_{{\rm I}}} \frac{{\rm d}R_{{\rm
             I}}  }{{\rm d} E_{{\rm ee}}} (E_{{\rm ee}})  \frac{{\rm d}
             E_{{\rm ee}}}{ \Delta E} \,,
\end{eqnarray}
where $r_{{\rm Na}}=0.153$ and $r_{{\rm I}}=0.847$ are the mass
fractions of Na and I respectively.

Since $v_{0} \gg v_{1}$ the differential event rate in the k-th energy bin
can be expanded in a Taylor series in $\cos \alpha$~\cite{ffg}:
\begin{equation} 
\frac{\Delta R}{\Delta E_{{\rm }}}(E_{{\rm k}}) \approx S_{{\rm 0, k}} 
           + S_{{\rm m, k}} \cos \alpha \,.
\end{equation}
The DAMA collaboration use a maximum likelihood method to separate the
time-independent background from the WIMP signal and have released the
resulting values of $S_{{\rm 0,k}}$ and $S_{{\rm m, k}}$, and the
errors on them, for each energy bin~\cite{newdama} (see Table 1).

\begin{table}
\begin{center}
\begin{tabular}{|c|c|c|}
Energy (keV) & $S_{{\rm 0,k}}$ (cpd/kg/keV) & $S_{{\rm m,k}}$ (cpd/kg/keV)  \\
2-3 & 0.54 $\pm$ 0.09 & 0.023 $\pm$ 0.006 \\
3-4 & 0.21 $\pm$ 0.05 & 0.013 $\pm$ 0.002 \\
4-5 & 0.08 $\pm$ 0.02 & 0.007 $\pm$ 0.001 \\
5-6 & 0.03 $\pm$ 0.01 & 0.003 $\pm$ 0.001
\end{tabular}
\end{center}
\caption[dama]{\label{dama} $S_{{\rm 0,k }} $ and $S_{{\rm m, k}}$
values obtained, by the DAMA collaboration, from a maximum likelihood
analysis of their combined data from four annual cycles.}
\end{table}

\section{Halo models}
\label{hm}
Since the resolution of numerical simulations is not yet large enough
to allow the numerical calculation of the annual modulation
signal~\cite{WD}, we have to use analytic models for the velocity
distribution of the dark matter halo. If the galactic halo contains
large amounts of substructure, as N-body simulations appear to
indicate~\cite{Nbody}, then the WIMP detection rate may be
dramatically altered if we are currently passing through a
clump of substructure.  The extent to which substructure is actually
present in the halo is currently a matter of debate however.

The effect of the halo model on the annual modulation signal can be
assessed most simply, with least recourse to the detector properties,
in terms of the dimensionless function $T(E)$~\footnote{The shape of
the energy spectrum measured also depends on the nuclear form factor
$F(q)$ however.}, as defined in Eq.~(\ref{tq}). The yearly averaged
value of $T(E)$ depends weakly on the halo model~\cite{kk} decreasing
with increasing recoil energy, with the decrease being most rapid for
small WIMP masses.  For each model we plot the annual variation in
$T(E)$:
\begin{equation}
\Delta T(E) = \frac{(T(E)_{{\rm max}}-T(E)_{{\rm
av}})}{ T(E)_{{\rm av}}} \,, 
\end{equation}
defined such that $\Delta T(E)$ is taken to be positive if $T(E)$ is
largest in June ($\alpha = 0$) as found by the DAMA collaboration, as
a function of recoil energy for a ${{\rm Ge}}^{76}$ detector and four
values of the WIMP mass: $m_{\chi}=30,50,100,200$ GeV. Values for other
monatomic detectors can be found by rescaling the x-axis by $m_{{\rm
A}}/(m_{{\rm A}} + m_{\chi})^{2}$.

\subsection{Asymmetric velocity distribution}
An extension of the Michie model can be used as a simple model
of velocity anisotropy~\cite{am1,Vergados}
\begin{eqnarray}
f({\bf v})&=& N \left[ \exp{ \left( -\frac{ v^2}{\sigma^2} \right)} 
          - \exp{ \left( - \frac{v_{{\rm esc}}}{\sigma^2} \right)}
          \right]   \nonumber \\
          &&  \times \exp{ \left( - \lambda \frac{ v_{\phi}^2 + 
          v_{{\rm z}}^2 }{\sigma^2} \right) } \,,
\end{eqnarray}
where $v_{{\rm esc}}$ is the escape velocity and $N$ is a
normalisation constant.  The deviation of the velocity distribution
from isotropic is parameterised by $\lambda$, the standard Maxwellian
halo, cut-off at the escape velocity, is recovered for
$\lambda=0$. Since the halo is formed by gravitational collapse the
deviation is most likely to be towards radial orbits, with $0 < \lambda
< 1$~\cite{am1}. We take $v_{{\rm esc}}= 650 {{\rm km s}}^{-1}$,
although the effect of variations in the value of the escape velocity
is small~\cite{am1,rot2}.

\begin{figure}[t]
\centering
\leavevmode\epsfysize=6.5cm \epsfbox{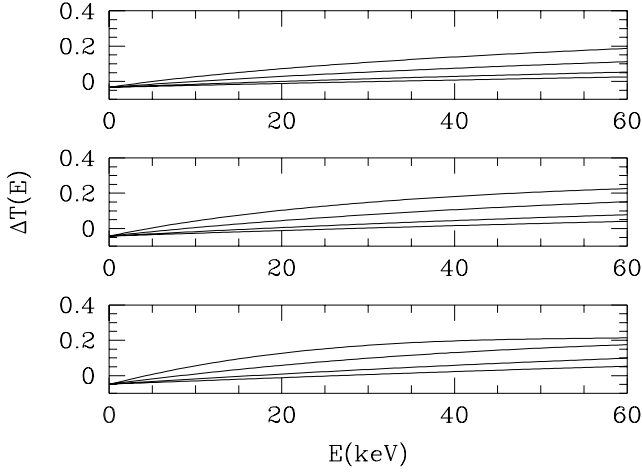}\\
\caption[fig1]{\label{fig1} The annual variation in $T(E)$, $\Delta T
(E)$, for halo models with asymmetric velocity distributions with
$\lambda=0,0.5,1$, from top to bottom row, for four values of the WIMP
mass $m_{\chi}=30,50,100, 200$ GeV, from bottom to top for a ${{\rm
Ge}}^{76}$ detector.}
\end{figure}

The annual variation in $T(E)$, $\Delta T(E)$, is plotted in
Fig.~\ref{fig1} for $\lambda=0,0.5,1$.  The mean value of $T(E)$ is
slightly higher for the asymmetric models than for the standard
Maxwellian halo model.

\subsection{Power-law halos}
Evans' family of axisymmetric distribution functions~\cite{evans} lead
to velocity distribution functions, in the rest frame of the galaxy, of
the form~\cite{kk}:
\begin{eqnarray}
f({\bf v})&=& ( A R^2 v_{\phi}^2 + B) 
           \frac{ \exp{ \left[ -2(v/v_{0})^2 \right] }}
           {(R^2 +R_{{\rm c}}^2)^2}
            \nonumber \\ && + \frac{ \exp{ \left[ -2(v/v_{0})^2
                \right] }}
             {R^2 +R_{{\rm c}}^2} \,,
\end{eqnarray}
with 
\begin{eqnarray}
A&=& \left( \frac{2}{\pi} \right)^{5/2} \frac{ (1-q^2)}{G q^2 v_{0}^3} \,,
B=\left( \frac{2}{\pi^5} \right)^{1/2} \frac{R_{{\rm c}}}{G q^2 v_{0}} \,,
 \nonumber \\
C&=& \frac{ 2 q^2 -1}{ 4 \pi G q^2 v_{0}} \,,
\end{eqnarray}
where $R_{{\rm c}}$ is the core radius, $R_{0}$ is the solar radius
and $q$ is a flattening parameter, which varies between 1, for
spherical halo, and $1/ \sqrt{2}=0.707$. Following Ref.~\cite{kk} we
take $R_{{\rm c}}=8.5$kpc and $R_{0}=7$kpc and explore the effect of
varying $q$. 

\begin{figure}
\centering
\leavevmode\epsfysize=6.5cm \epsfbox{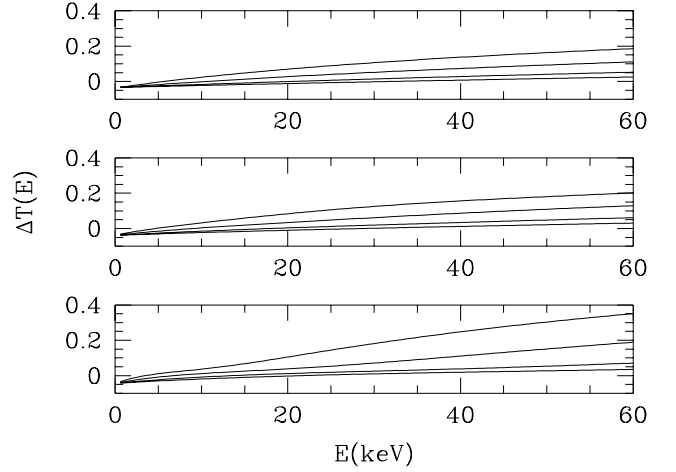}\\
\caption[fig2]{\label{fig2} The annual variation in $T(E)$, $\Delta
T(E)$, for power-law halos with $q=1,0.85, 0.707$, from top to bottom
row, for four values of the WIMP mass $m_{\chi}=30,50,100, 200$ GeV,
from bottom to top for a ${{\rm Ge}}^{76}$ detector.}
\end{figure}

The annual variation in $T(E)$, $\Delta T(E)$, is plotted in
Fig.~\ref{fig2} for $q=1,0.85,0.707$. The mean value of $T(E)$ is
slightly lower for the flattened halos than for the Maxwellian
halo. The change in the annual variation due to flattening is larger
than that due to an asymmetric velocity distribution, particularly so
for large recoil energies and small WIMP masses.

\subsection{Bulk rotation}
\begin{figure}[t]
\centering
\leavevmode\epsfysize=6.5cm \epsfbox{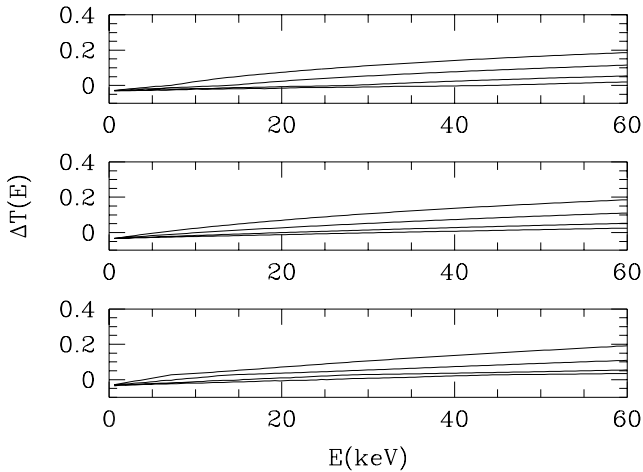}\\
\caption[fig3]{\label{fig3} The annual variation in $T(E)$, $\Delta
T(E)$, for rotating Maxwellian halo models with $a_{{\rm
rot}}=0.36,0.5,0.64$, from top to bottom row, for four values of the
WIMP mass $m_{\chi}=30,50,100, 200$ GeV, from bottom to top for a
${{\rm Ge}}^{76}$ detector.}
\end{figure}

Halo models with bulk rotation can be constructed by taking linear
combinations of the velocity distribution function~\cite{kk}:
\begin{equation}
f_{{\rm rot}}({\bf v}) = a_{{\rm rot}} f_{+}({{\bf v}}) + 
               (1-a_{{\rm rot}})f_{-}({{\bf v}}) \,,
\end{equation}
where
\begin{equation}
f_{+}({{\bf v}}) = \left\{ \begin{array}{ll}
                   f({\bf v}) \quad v_{\phi} > 0  \\
                    0 \,\, \qquad  v_{\phi} < 0 \end{array} \right.
\end{equation}
\begin{equation}
f_{-}({{\bf v}}) = \left\{ \begin{array}{ll}
                   0   \,\, \qquad   v_{\phi} > 0  \\
                   f({\bf v})    \quad   v_{\phi} < 0 \end{array} \right.
\end{equation}
and $a_{{\rm rot}}$ is related to the dimensionless galactic angular
momentum, $\lambda_{{\rm rot}}$: $\lambda_{{\rm rot}}= 0.36 | a_{{\rm
rot}} -0.5|$. Numerical studies of galaxy formation find that
$|\lambda_{{\rm rot}}| < 0.05$ \cite{gf}, corresponding to $ 0.36 <
a_{{\rm rot}} < 0.64$. A non-rotating halo has $a_{{\rm rot}}=0.5$,
whilst a counter-rotating (co-rotating) has $a_{{\rm rot}}<0.5$
($a_{{\rm rot}}>0.5$).

The annual variation in $T(E)$, $\Delta T(E)$, is plotted in
Fig.~\ref{fig3} for Maxwellian halos with $a_{{\rm
rot}}=0.36,0.5,0.64$.  The mean value for counter(co)-rotation is
lower (higher) than for the non-rotating Maxwellian halo at small
recoil energies, and higher (lower) at large recoil energies. The
change in the annual variation, which is suppressed (enlarged) for
counter(co)-rotation, is largest for small recoil energies.

\vspace{5mm}

For each of the halo models studied  
the
annual variation is negative for small recoil energies ($E \lesssim 4$
keV for $M_{\chi}=30$GeV, $E \lesssim 30$ keV for $M_{\chi}=200$ GeV)
increasing as the recoil energy increases. For fixed recoil energy the
annual variation is largest for small WIMP masses.

\section{Analysis of the DAMA data}
\begin{figure}
\centering
\includegraphics[angle=270,width=0.45\textwidth]{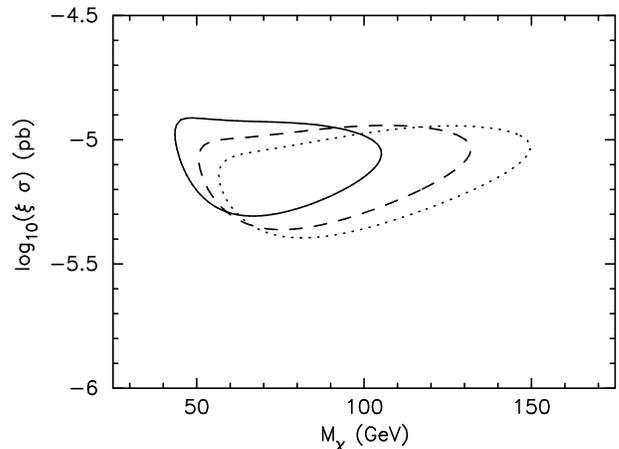}
\caption[fig4]{\label{fig4} The $\kappa=35$ contour, delineating the
region of $m_{\chi}- \xi \sigma$ parameter space compatible with
the DAMA annual modulation signal, for asymmetric halo models with
$\lambda=0,0.5,1$ (solid, dashed and dotted lines respectively).}
\end{figure}

Brhlik and Roszkowski~\cite{br} have devised a technique for comparing
the experimental results released by DAMA with theoretical predictions
for the annual modulation signal, in the absence of detailed
information about the experimental set-up, such as the efficiency of
each NaI crystal. They define a function $\kappa$:
\begin{equation}
\kappa= \Sigma_k \frac{ (S_{{\rm 0,k}}^{{\rm th}} - S_{{\rm 0,k}}^{{\rm 
               exp}})^2}{\sigma_{{\rm 0,k}}^2} + 
           \Sigma_k \frac{ (S_{{\rm m,k}}^{{\rm th}} - S_{{\rm m,k}}^{{\rm 
               exp}})^2}{\sigma_{{\rm m,k}}^2} \,,
\end{equation}
where the experimental errors on the time dependent and independent
parts of the signal, $\sigma_{{\rm m,k}}$ and $\sigma_{{\rm 0,k}}$
respectively, serve as weights. The contour, in the $m_{\chi}-\xi
\sigma_{{\rm p}}$ plane, $\kappa=35$ agrees reasonably well with the
DAMA collaborations $3\sigma$ contour.  Whilst this approach does not
give accurate confidence limits on $m_{\chi}$ and $\xi \sigma_{{\rm
p}}$ it does illustrate the qualitative effect of varying the
properties of the halo model on the values of the WIMP parameters
obtained from the data. A full likelihood analysis of the DAMA data
has been carried out for models with bulk rotation in
Ref.~\cite{rot2}, allowing us to check the reliability of Brhlik and
Roszkowski's technique.

Given the experimental difficulties of extracting a small annual
variation from, possibly time dependent, backgrounds, concerns have
been expressed about the interpretation of the earlier (1 and 2 year)
DAMA data as evidence for a WIMP signal~\cite{concern}. Furthermore
the Cryogenic Dark Matter Search (CDMS) collaboration have recently
released limits on the WIMP-Nucleon cross section which exclude, at
85$\%$ confidence, the entire DAMA $3\sigma$ allowed
region~\cite{CDMS}. The DAMA collaboration have, however, performed a
thorough analysis of the various sources of possible systematic
errors~\cite{damase} and, since the DAMA and CDMS experiments use
different target nuclei, assumptions are required to perform a direct
comparison.  In any case even if the interpretation of the DAMA annual
modulation signal as WIMP scattering is eventually found to be
erroneous, our results will still indicate the qualitative effect of
these non-standard halo models on the analysis of a WIMP annual
modulation signal.

\begin{figure}
\centering
\includegraphics[angle=270,width=0.45\textwidth]{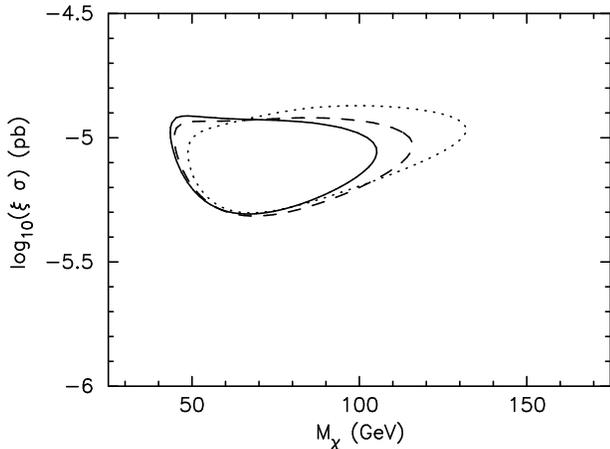}
\caption[fig5]{\label{fig5} 
The $\kappa=35$ contour for flattened halo models with $q=1.0,0.85,0.707$
(solid, dashed and dotted lines respectively).}
\end{figure}

\begin{figure}
\centering
\includegraphics[angle=270,width=0.45\textwidth]{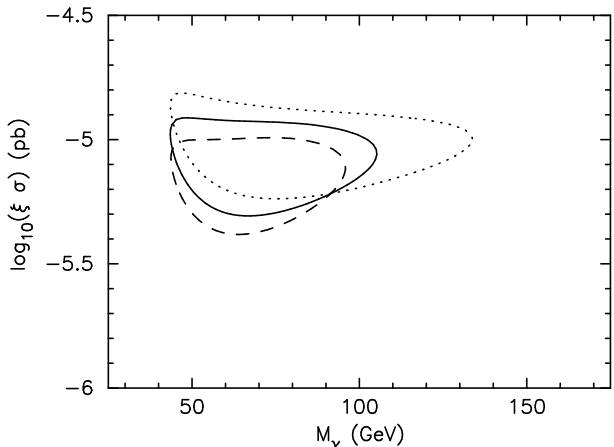}
\caption[fig6]{\label{fig6} 
The $\kappa=35$ contour for the
standard halo model with $a_{{\rm rot}}=0.36,0.5,0.64$ (dashed, solid and 
dotted lines respectively). The non-rotating halo corresponds to 
$a_{{\rm rot}}=0.5$.}
\end{figure}

In Figs.~\ref{fig4}-~\ref{fig6} we plot contours of $\kappa=35$,
delineating the region of $m_{\chi}$-$\xi \sigma$ parameter space
compatible with the DAMA annual modulation signal, for asymmetric,
flattened and rotating halo models respectively.  In Fig.~\ref{fig4}
we see that as the asymmetry of the velocity distribution is increased
the allowed region is enlarged and moves to larger masses and slightly
smaller interaction cross sections. We saw in Fig.~\ref{fig2} that the
change in the annual modulation signal for the flattened halo model is
larger than that for the asymmetry velocity
distribution. Consequentially the change in the allowed region for the
flattened halo model is smaller, extending to larger masses and
slightly larger interaction cross sections.  In Fig.~\ref{fig6} we can
see that counter-rotation ($a_{{\rm rot}}<0.5$) contracts the allowed
region and shifts it to smaller cross sections whilst co-rotation
($a_{{\rm rot}}>0.5$) expands it and shifts it to larger cross
sections. This is in good agreement with the results of
Ref.~\cite{rot2} on the effect of bulk rotation, which were found via
a full likelihood analysis of the DAMA data.

\section{Conclusions}
In this paper we have examined the effect of an asymmetric halo
velocity distribution, halo flattening and bulk halo rotation on the
WIMP annual modulation signal. Whilst for each of the halo models the
change in the mean signal is small, $< 10 \%$ for the range of recoil
energies probed by the DAMA experiment, the amplitude of the annual
modulation can change significantly. The magnitude of the change
depends on the WIMP mass and recoil energy, as well the nature of the
deviation of the halo model from the standard Maxwellian halo. With
bulk rotation the change in the annual variation, relative to the
Maxwellian halo model, is largest for small recoil energies, whilst
halo flattening produces the largest change for larger recoil
energies.

The range of WIMP masses consistent with the DAMA annual modulation
signal is enlarged significantly, to roughly $ 30 < m_{\chi}< 150$ GeV
at $3\sigma$, for each of the models considered. The shape of the
allowed region in the $m_{\chi}-\xi \sigma_{{\rm p}}$ plane is
different for each model however. This indicates that the
uncertainties in halo modelling have a significant effect on the WIMP
mass determined from the annual modulation signal. Therefore more
sophisticated halo models (such as those in Ref.~\cite{newevans}) need
to be developed and used in the analysis of annual modulation data.

\acknowledgments
The author is supported by PPARC and acknowledges the use of the Starlink 
facilities at QMW. The author would also like to thank J. D. Vergados for 
useful discussions.

\appendix
\section{Interaction Cross Section}
The differential cross section for neutralino scattering off a target nucleus
is dominated by the scalar and axial terms~\cite{gw,jkg,br}:
\begin{equation}
\frac{{\rm d} \sigma}{{\rm d} {\bf q}^2}= \frac{{\rm d} \sigma^{{\rm scalar}}}
         {{\rm d} {\bf q}^2} + \frac{{\rm d} \sigma^{{\rm axial}}}
          {{\rm d} {\bf q}^2} \,,
\end{equation}
where ${\bf q}=(m_{{\rm A}} m_{\chi})/(m_{{\rm A}} + m_{\chi}) {\bf
v}$ is the momentum transfered.

The scalar differential cross section, which arises due to Higgs boson
and squark exchange, is given by~\cite{jkg}
\begin{equation}
\frac{{\rm d} \sigma^{{\rm scalar}}}
         {{\rm d} {\bf q}^2}= \frac{1}{\pi v^2} \left[ Z f_{{\rm p}} +
            (A-Z) f_{{\rm n}} \right]^2 F^2(q) \,,
\end{equation}
where $f_{{\rm p}}$ and $f_{{\rm n}}$ are the effective neutralino
couplings to the proton and the neutron respectively and $F(q)$ is the
scalar nuclear form factor. The form factor for Na is usually taken to
be equal to unity, whilst for I the Saxon-Woods form
factor~\cite{swff}
\begin{equation}
F(q)=\frac{3 j_{1}(qR_{1})}{q R_{1}} \exp{\left[ -  (qs)^2 /2 \right]} \,,
\end{equation}
where $R_{1}=\sqrt{R_{{\rm A}}^2 -5 s^2}$, $R_{{\rm A}}= A^{1/3}
\times 1.2$fm and $s=1$fm, is used.

The axial differential cross section, which arises due to $Z_{0}$ and
squark exchange, is given by~\cite{jkg}
\begin{equation}
\frac{{\rm d} \sigma^{{\rm axial}}}
          {{\rm d} {\bf q}^2} = \frac{8}{ \pi v^2}\Lambda^2 J(J+1)S(q) \,, 
\end{equation}
where $J$ is the total angular momentum, $S(q)$ is the spin form
factor and $\Lambda$ depends on the axial couplings of the neutralino
to the quarks (see Ref.~\cite{jkg} for more details and explicit
expressions).

The differential event rate for a given detector can then be expressed
as~\cite{jkg}:
\begin{eqnarray}
\frac{{\rm d}R}{{\rm d} E}&=& \frac{ 4 }{\pi^{3/2}} \frac{\rho_{\chi}}
             {m_{\chi}} T(E) \\ \nonumber
            &&  \times \{ \left[ Zf_{{\rm p}} + (A-Z)f_{{\rm n}} \right]^2 
            F^2(q)
             + 8 \Lambda^2 J(J+1)S(q) \}  \,.
\end{eqnarray}


\begin{thebibliography}{99}
\bibitem{ash} K. M. Ashman, Publ. Astron. Soc. Pac., 104, 1109 (1992); 
            C. J. Kochanek, 
               Astophys. J. {\bf 445}, 559 (1995).
\bibitem{pss} J. R. Primack, B. Sadoulet and D. Seckel, Ann. Rev. 
               Nucl. Part. Sci., B38, 751 (1988).
\bibitem{ck} C. J. Copi, J. Heo and L. M. Krauss, Phys. Lett. {\bf B461},
             43 (1999).
\bibitem{am1} A. K. Drukier, K. Freese and D. N. Spergel, Phys. Rev. 
            D {\bf 33}, 3495 (1986). 
\bibitem{am2} K. Freese. Phys.
              Rev. D {\bf 37}, 3388 (1988). 
\bibitem{dama} R. Bernabei et. al. Phys. Lett. {\bf B389}, 757 (1996); 
               ibid {\bf B408}, 439 (1997); ibid {\bf B424}, 195 (1998); 
                 ibid {\bf B450}, 448 (1999). 
\bibitem{newdama} R. Bernabei et. al. Phys. Lett. {\bf B480}, 23 (2000).

\bibitem{rot1} F. Donato, N. Fornengo and S. Scopel, Astropart. Phys. {\bf 9},
              247 (1998).
\bibitem{rot2} P. Belli et. al. Phys. Rev. D {\bf 61}, 023512 (2000).
\bibitem{br} M. Brhlik and L. Roskzkowski, Phys. Lett. {\bf B464}, 303 (1999).
\bibitem{vrange} P. J. T. Leonard and S. Tremaine, Astrophys. J. {\bf 353},
             486 (1990); K. M. Cudworth, Astron. J. {\bf 99}, 590 (1990);
            C. S. Kochanek, Astrophys. J. {\bf 457}, 228
             (1996).
\bibitem{alcock95} C. Alcock et. al. Astrophys. J., {\bf 449}, 28 (1995). 
\bibitem{Sackett} P. D. Sackett et. al. Astrophys. J., {\bf 436}, 629 (1994).
\bibitem{evans} N. W. Evans, Mon. Not. Roy. Astron. Soc., {\bf 260}, 
                191 (1993); ibid. {\bf 267}, 333 (1994).
\bibitem{kk} M. Kamionkowski and A. Kinkhabwala, Phys. Rev. D {\bf 57}, 3256
            (1998).
 \bibitem{Vergados} J. D. Vergados, Phys. Rev. Lett. {\bf 83}, 3597 (1999);
                   Phys. Rev. D {\bf 62}, 023519 (2000).
\bibitem{UK} P. Ullio and M. Kamionkowski, hep-ph/0006183.
\bibitem{griest} K. Griest, Phys. Rev. D {\bf 37}, 2703, (1988).
\bibitem{jkg} G. Jungman, M. Kamionkowski and K. Griest, Phys. Rep. 267, 
               195 (1996).
\bibitem{q} K. Fushimi et. al. Phys. Rev. C {\bf 47}, R245 (1993);
            G. J. Davies et. al. Phys. Lett. {\bf B322}, 159 (1994);
            P. F. Smith et. al. Phys. Lett. {\bf B379}, 299 (1996).

\bibitem{ffg} K. Freese, J. Frieman and A. Gould, Phys. Rev. D {\bf 37},
              3388 (1988).
\bibitem{WD} L. M. Widrow and  J. Dubinski Astrop. J., {\bf 504}, 12  (1998). 

\bibitem{Nbody} J. F. Navarro, C. S. Frenk and S. D. M. White, Astophys.
             J. {\bf 462}, 563 (1996); B. Moore et. al. Mon. Not. R. 
            Astron. Soc. {\bf 310} (1999), 1147; A. V. Kravtsov et. al. 
            Astrophys. J. {\bf 502}, 48 (1998).
\bibitem{gf} J. Barnes and G. Efstathiou, Astrophys. J. {\bf 319}, 575
            (1987); M. S. Warren, P. J. Quinn, J. K. Salomon and
            W. H. Zurek, Astrophys. J. {\bf 399}, 405 (1992); S. Cole
            and C. Lacey, Mon. Not. R. Astron. Soc. {\bf 281}, 716 (1996).
\bibitem{concern} G. Gerbier, J. Mallet, L. Mosca and C. Tao, 
                astro-ph/9710181; astro-ph/9902194.
\bibitem{CDMS} CDMS Collaboration, Nucl. Instrum. Meth. {\bf A444}
               345  (2000); Phys. Rev. Lett. {\bf 84}, 5699 (2000).
\bibitem{damase} R. Bernabei et. al. preprint ROM2F/2000-15, to
           appear in the proceedings of the International Workshop 
          DM2000, Marina del Rey.
\bibitem{newevans} N. W. Evans, C. M. Carollo and P. T. de Zeeuw, 
           astro-ph/0008156.
\bibitem{gw} M. Goodman and E. Witten, Phys. Rev. D {\bf 31}, 3059 (1985).
\bibitem{swff} J. Engel, Phys. Lett. {\bf B264}, 114 (1991).

\end{thebibliography}
\end{document}